\documentclass[10pt]{iopart}
\usepackage{graphicx,iopams,bm}
\begin{document}
\jl{12}
\submitted

\paper[Beam customization in the pencil beam splitting algorithm]{Modeling of beam customization devices in the pencil beam splitting algorithm for heavy charged particle radiotherapy}

\author{Nobuyuki~Kanematsu$^{1,2}$}
\address{$^1$\ Department of Accelerator and Medical Physics, Research Center for Charged Particle Therapy, National Institute of Radiological Sciences, 4-9-1 Anagawa, Inage-ku, Chiba 263-8555, Japan}
\address{$^2$\ Department of Quantum Science and Energy Engineering, School of Engineering, Tohoku University, 6-6 Aramaki Aza Aoba, Aoba-ku, Sendai 980-8579, Japan}

\eads{nkanemat@nirs.go.jp}

\begin{abstract}
A broad-beam-delivery system for heavy-charged-particle radiotherapy often employs multiple collimators and a range-compensating filter, which potentially offer complex beam customization.
In treatment planning, it is however difficult for a conventional pencil-beam algorithm to deal with these structures due to beam-size growth during transport.
This study aims to resolve the problem with a novel computational model.
The pencil beams are initially defined at the range compensating filter with angular-acceptance correction for the upstream collimators followed by the range compensation effects.  
They are individually transported with possible splitting near the downstream collimator edges to deal with its fine structure.
The dose distribution for a carbon-ion beam was calculated and compared with existing experimental data.
The penumbra sizes of various collimator edges agreed between them to a submillimeter level. 
This beam-customization model will complete an accurate and efficient dose-calculation algorithm for treatment planning with heavy charged particles.
\end{abstract}


\section{Introduction}

For heavy charged particle radiotherapy with protons and ions, broad-beam delivery methods (Coutrakon \etal 1991, Kanai \etal 1999) are mature technologies with persistent advantages of simplicity and robustness over emerging technologies of pencil-beam scanning methods (Lambert \etal 2005).
For a broad-beam system, a variety of volumetrically enlarged standard beams are prepared, among which an optimum one is applied to a given target.
Target-specific customization is usually made with x-jaw, y-jaw, and multileaf collimators (XJC, YJC, and MLC) and custom-made accessories such as a patient collimator (PTC) and a range-compensating filter (RCF).
While the downstream collimators form sharp field edges, the upstream collimators, which are mainly for radiation-protection purposes, form gentle field edges.
Their combination will be useful for field patching techniques to form an irregular field with gently joining beams for improved robustness (Li \etal 2007).

In treatment planning, a variety of pencil-beam (PB) algorithms are used for dose calculation (Hong \etal 1996, Kanematsu \etal 2006) despite intrinsic difficulty with the pencil beams that develop to overreach lateral heterogeneity (Goitein 1978, Petti 1992, Kohno \etal 2004).
For electron radiotherapy, the phase-space theory was rigorously applied to resolve the problem by periodical redefinition of ensemble of minimized pencil beams in the PB-redefinition algorithm (Shiu and Hogstrom 1991).
The principle of PB redefinition was applied to heavy charged particles to address the effects of multiple collimators in the monochromatic PB approximation (Kanematsu \etal 2008b).
However, its rigorous application to a heterogeneous system requires polychromatic energy spectra, which would be computationally demanding for heavy charged particles with sharp Bragg peaks.
It will be thus difficult to cope with range compensation or patient heterogeneity in that approach.

Recently, Kanematsu \etal (2009) proposed an alternative approach, the PB-splitting algorithm, where monochromatic pencil beams dynamically split into smaller ones near a lateral density interface.
Automatically, fine pencil beams are densely arranged only where they are necessary while otherwise large pencil beams are sparsely arranged for efficient dose calculation.
In conjunction with the grid-dose-spreading convolution (Kanematsu \etal 2008a), the PB-splitting algorithm demonstrated feasibility of accurate patient dose calculation while minimizing the impact of recursive beam multiplication (Kanematsu 2011).

In this study, we further extend the PB-splitting approach to beam-customization devices to deal with  their physical structures accurately and efficiently and to complete a consistent algorithmic framework for dose calculation in treatment planning.
In the following sections, we define the model elements that were mostly diverted from previous studies, construct a novel and original beam-customization model, and examine its validity for a test-beam experiment.

\section{Materials and methods}

\subsection{Model elements}

\subsubsection{Beam source and beam's eye view}

A beam source is defined as the best approximate point from which radiating particles will have the same fluence reduction with distance.
The formulation differs among beam-spreading methods and often between the transverse $x$ and $y$ axes, {\it i.e.}, at height $z_{{\rm S}x}$ for $x$ and $z_{{\rm S}y}$ for $y$.
The particles incoming to a point in the field, which is normally the isocenter, are projected back onto the $x$ and $y$ source planes to define rms source sizes $\sigma_{{\rm S}x}$ and $\sigma_{{\rm S}y}$ in the Gaussian approximation.
Although a range-modulated beam should be ideally subdivided into energy components of different source heights and sizes, it is approximately represented by a single component of average behavior in this study. 

A beam's-eye-view (BEV) image is defined as an $n \times m$ matrix of $\delta \times \delta$ square-sized pixels starting at $(x_1, y_1)$ on the isocenter plane. 
For BEV pixel $i j$, pixel position $(x_j, y_i)$ and the line connecting to the $x$ and $y$ sources are defined as
\begin{eqnarray}
\fl
\left(x_j,y_i\right) = \left(x_1,y_1\right) + \left(j-1,i-1\right) \delta,\qquad
\cases{x(x_j, z) = \frac{z_{{\rm S}x}-z}{z_{{\rm S}x}} x_j\\ y(y_i, z) = \frac{z_{{\rm S}y}-z}{z_{{\rm S}y}} y_i.}\label{eq:1}
\end{eqnarray}

\subsubsection{Collimator}

Following the thick-collimator model (Kanematsu \etal 2006), two identical apertures on the top and bottom faces are associated with every collimator, which are modeled as two-dimensional bitmaps. 
Matrix ${\bm T}_a$ describes aperture $a$ with $n_a \times m_a$ elements of transmission ${T_a}_{i_a j_a} =$ 1 (transmit) or 0 (block) for $i_a \in [1,n_a]$ and $j_a \in [1, m_a]$.
The pixel-$i_a j_a$ position is given by
\begin{eqnarray} 
({x_a}_{j_a}, {y_a}_{i_a}) = ({x_a}_1, {y_a}_1 ) + (j_a-1, i_a-1) \delta_a, \label{eq:2}
\end{eqnarray} 
where $({x_a}_1, {y_a}_1)$ and $\delta_a$ are the first pixel position and the square pixel size of the bitmap image.
For arbitrary point $(x_a,y_a)$ on the aperture plane, intersecting pixel $i_a j_a$ is determined with the nearest integer function $\lfloor\,\rceil$ as
\begin{eqnarray}
j_a = \left\lfloor \frac{x_a-{x_a}_1}{\delta_a} \right\rceil+1,\qquad
i_a = \left\lfloor \frac{y_a-{y_a}_1}{\delta_a} \right\rceil+1, \label{eq:3}
\end{eqnarray}
and the distance to the nearest aperture edge, 
\begin{eqnarray}
{d_a}_{i_a j_a} = \min_{\{i' j' \mid {T_a}_{i'_a j'_a} \neq {T_a}_{i_a j_a}\}} \left(\delta_a \sqrt{(i'_a-i_a)^2+(j'_a-j_a)^2}\right), \label{eq:4}
\end{eqnarray}
is quickly referenced from the distance map filled by the distance-transform algorithm (Borgefors 1986).

\subsubsection{Range compensating filter}

A RCF made of a tissue-like material of effective density $\rho_{\rm C}$ is similarly described by an $n_{\rm C} \times m_{\rm C}$ matrix of range shifts ${\bm S}_{\rm C}$, first pixel position $({x_{\rm C}}_1, {y_{\rm C}}_1)$, and pixel size $\delta_{\rm C}$. 
In this study, we deal with a single RCF of a flat downstream face at height $z_{\rm C}$. 
The stopping and scattering effects of the RCF are approximated by a local interaction at the midpoint of the beam path in the structure (Gottschalk \etal 1993), {\it i.e.}, at height $z = z_{\rm C} + 0.5\, {S_{\rm C}}_{i_{\rm C} j_{\rm C}}/\rho_{\rm C}$ for RCF pixel $i_{\rm C} j_{\rm C}$.

\subsubsection{Pencil beam}

Following the original PB-splitting algorithm (Kanematsu \etal 2009), the present PB model is based on the Fermi-Eyges theory (Eyges 1948) for stopping and scattering (Kanematsu 2009, Gottschalk 2010) excluding hard interactions that are implicitly included in the depth--dose curve.

A Gaussian pencil beam is characterized by position $\vec{r}$, direction $\vec{v}$, number of particles $n$, residual range $R$, and phase-space variances of the projected angle $\theta$ and transverse displacement $t$, which develop in a tissue-like medium by step $\Delta s$ as
\begin{eqnarray}\fl
\Delta \vec{r} = \vec{v}\, \Delta s, \qquad \Delta \vec{v} = \vec{0},\qquad 
\Delta n = 0, \qquad \Delta R = -\rho\, \Delta s, \label{eq:5} \\ \fl 
\Delta \overline{\theta^2} = \frac{1.00}{1000} q^{-0.16} \left(\frac{m}{m_p}\right)^{-0.92} \ln\frac{R}{R+\Delta R},\label{eq:6} \\ \fl
\Delta \overline{\theta t} = \left(\overline{\theta^2}+\frac{1}{2} \Delta \overline{\theta^2}\right) \Delta s,\qquad
\Delta \overline{t^2} = \left[2\, \overline{\theta t}+\left(\overline{\theta^2}+\frac{1}{3} \Delta \overline{\theta^2}\right) \Delta s \right] \Delta s, \label{eq:7}
\end{eqnarray}
where $\rho$ is the stopping-power ratio of the medium to water (Kanematsu \etal 2003) and $m/m_p$ and $q$ are the particle mass and charge in units of those of a proton. 

To limit excessive beam multiplication, pencil beams subject to splitting should have sufficient particles, {\it i.e.}, $n/n_0 > \kappa_{\rm n}$, where $n_0$ is the number for the original beam and $\kappa_{\rm n}$ is a cutoff.
When a pencil beam of rms size $\sigma_t = \surd(\overline{t^2})$ spreads beyond the lateral density interface at distance $d_{\rm int}$ from the beam center, it splits into $M \times M$ daughter beams downsized by factor $\sigma_M$ as
\begin{eqnarray}
\fl
M = \cases{
2 & for \quad $\sigma_2\, \kappa_{\rm d}\, \sigma_t < d_{\rm int} \le \kappa_{\rm d}\,\sigma_t$\\
3 & for \quad $\sigma_3\, \kappa_{\rm d}\, \sigma_t < d_{\rm int} \le \sigma_2\, \kappa_{\rm d}\,\sigma_t$\\
4 & for \quad $d_{\rm int} \le \sigma_3\, \kappa_{\rm d}\,\sigma_t$},\qquad
\left(\begin{array}{c}\sigma_2\\\sigma_3\\\sigma_4\end{array}\right) = \left(\begin{array}{c}\sqrt{3}/2\\1/\sqrt{2}\\1/2\end{array}\right), \label{eq:8}
\end{eqnarray}
where $\kappa_{\rm d}$ is a parameter that limits the fraction of overreaching particles.
With respect to the mother beam, daughter $\alpha \beta$ ($\alpha, \beta \in [1,M]$) is downscaled, displaced, redirected, and downsized while conserving focal distance $\overline{t^2}/\overline{\theta t}$ and local mean square angle $\overline{\theta^2}-\overline{\theta t}^2/\overline{t^2}$,
as
\begin{eqnarray}\fl
n_{\alpha \beta} = {f_M}_\alpha {f_M}_\beta \, n,\qquad
\left(\begin{array}{c}{\bm f}_2\\ {\bm f}_3\\ {\bm f}_4\end{array}\right) = \left(\begin{array}{cccc} 1/2 & 1/2 \\ 1/4 & 1/2 & 1/4 \\ 1/8 & 3/8 & 3/8 & 1/8 \end{array}\right),\label{eq:9}\\ \fl
\vec{r}_{\alpha \beta} = \vec{r} + \Delta \vec{r}_{\alpha \beta},\qquad
\vec{v}_{\alpha \beta} = 
\left| \left(\overline{t^2}/\overline{\theta t}\right) \vec{v} + \Delta \vec{r}_{\alpha \beta} \right|^{-1}
\left[\left(\overline{t^2}/\overline{\theta t}\right) \vec{v} + \Delta \vec{r}_{\alpha \beta} \right],\label{eq:10}\\ \fl
\Delta \vec{r}_{\alpha \beta} = \sigma_t \left( {\mu_M}_\alpha \vec{e}_t + {\mu_M}_\beta \vec{e}_u \right),\qquad
\left(\begin{array}{c}{\bm \mu}_2\\ {\bm \mu}_3\\ {\bm \mu}_4\end{array}\right) = \left(\begin{array}{cccc} -1/2 & 1/2\\ -1 & 0 & 1 \\ -3/2 & -1/2 & 1/2 & 3/2 \end{array}\right),\label{eq:11}\\ \fl
\overline{t^2}_{\alpha \beta} = \sigma_M^2\,\overline{t^2},\qquad
\overline{\theta t}_{\alpha\beta} = \sigma_M^2\,\overline{\theta t},\qquad
\overline{\theta^2}_{\alpha\beta} = \overline{\theta^2}-(1-\sigma_M^2)\,\overline{\theta t}^2/\overline{t^2}, \label{eq:12}
\end{eqnarray}
where ${f_M}_\alpha {f_M}_\beta$ is the share of daughter $\alpha\beta$ and ${\mu_M}_\alpha$ and ${\mu_M}_\beta$ are the displacement factors for transverse directions $\vec{e}_t \approx \vec{e}_x$ and $\vec{e}_u \approx \vec{e}_y$. 

\subsection{The beam-delivery system model}

\subsubsection{Pencil-beam generation}

For every BEV pixel $i j$, pencil beam $b$ is placed and the PB parameters are defined at the effective interaction point in the RCF as
\begin{eqnarray}\fl
\vec{r}_b = \vec{r_0}_b = \left(x(x_j,{z_0}_{i j}), y(y_i,{z_0}_{i j}), {z_0}_{i j}\right), \qquad
{z_0}_{i j} = z_{\rm C} + 0.5\,{S_{\rm C}}_{i_{\rm C} j_{\rm C}}/\rho_{\rm C} \label{eq:13}
\\ \fl
\vec{v}_b = -\left[\left(\frac{x_j}{z_{{\rm S}x}}\right)^2+\left(\frac{y_i}{z_{{\rm S}y}}\right)^2+ 1\right]^{-1/2}\left(\frac{x_j}{z_{{\rm S}x}},\frac{y_i}{z_{{\rm S}y}}, 1\right), \label{eq:14}
\\ \fl
n_b = {n_0}_b = {\Phi_0}_{i j}\, \delta^2, \qquad R = R_0, \label{eq:15}
\\ \fl
\overline{\theta^2}_b = \frac{1}{2} \left(\frac{\sigma_{{\rm S}x}}{z_{{\rm S}x}-{z_0}_{i j}}\right)^2 + \frac{1}{2} \left(\frac{\sigma_{{\rm S}y}}{z_{{\rm S}y}-{z_0}_{i j}}\right)^2,
\label{eq:16}
\\ \fl
\overline{\theta t}_b = \frac{\overline{t^2}}{\sqrt{z_{{\rm S}x}-{z_0}_{i j}}\sqrt{z_{{\rm S}y}-{z_0}_{i j}}},\qquad \overline{t^2} = \frac{z_{{\rm S}x}-{z_0}_{i j}}{z_{{\rm S}x}} \frac{z_{{\rm S}y}-{z_0}_{i j}}{z_{{\rm S}y}}
\frac{\delta^2}{12}, \label{eq:17}
\end{eqnarray}
where RCF intersection pixel $i_{\rm C} j_{\rm C}$ is determined in analogy with \eref{eq:1} and \eref{eq:3} and open-field fluence ${\Phi_0}_{i j}$ and range $R_0$ are usually given by measurement.

\subsubsection{Upstream collimation}

\begin{figure}
\begin{indented}
\item[] \includegraphics[width=8.5 cm]{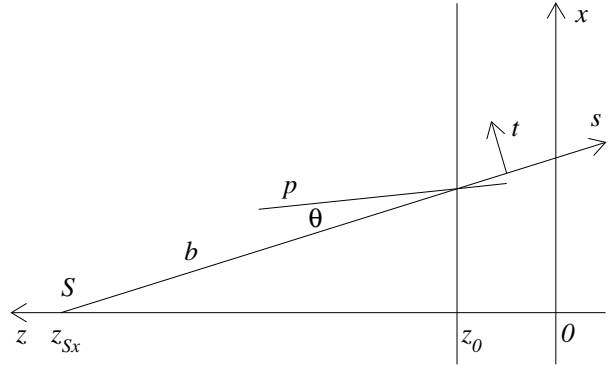}
\end{indented}
\caption{Definition of geometrical parameters (angle $\theta$ and axes $x$, $z$, $s$, and $t$) in the $z$--$x$ view, where symbols {\it S}, {\it b}, and {\it p} indicate the source, a pencil beam, and a particle, respectively.}
\label{fig:1}
\end{figure}

The upstream collimators restrict angular acceptance of particles incoming to each PB origin.
A pencil beam will be fully blocked when it is far away from any one of the apertures or fully transmitted when it is in the middle of all of them, or
\begin{eqnarray}
\fl T_b = \cases{0 & for $\quad \left\{^\exists a \mid (z_a > z_0) ; \left({T_a}_{i_a j_a} = 0\right) \wedge \left[ {d_a}_{i_a j_a} > 3\,{\sigma_\theta}_b (z_a-z_0) \right] \right\}$\\
1 & for \quad $\left\{ ^\forall a \mid (z_a > z_0) ; \left({T_a}_{i_a j_a} = 1\right) \wedge \left[ {d_a}_{i_a j_a} > 3\,{\sigma_\theta}_b (z_a-z_0) \right] \right\}$,} \label{eq:18}
\end{eqnarray}
where $T_b$ is the transmission factor of beam $b$, $i_a j_a$ is the beam intersection pixel of aperture $a$, and factor $3$ to rms projected angle ${\sigma_\theta}_b = \surd(\overline{\theta^2}_b)$ secures three standard deviations for edge distance ${d_a}_{i_a j_a}$. 
For partial transmission, we calculate the geometrical acceptance of particles incoming to the PB origin.
As shown in \fref{fig:1}, with small orthogonal angles $\theta$ and $\phi$ about the PB axis, constituent-particle direction $\vec{v}_p$ is defined as
\begin{eqnarray}
\vec{v}_p = \left({v_p}_x,{v_p}_y,{v_p}_z\right) \approx \vec{v}_b + \tan \theta\ \vec{e}_x + \tan \phi\ \vec{e}_y, \label{eq:19}
\end{eqnarray}
which translates into geometrical line $(\theta$,$\phi)$, or
\begin{eqnarray}
\cases{x(z) = x_0 + (z-z_0)\,{{v_p}_x}/{{v_p}_z}\\y(z) = y_0 + (z-z_0)\,{{v_p}_y}/{{v_p}_z}}. \label{eq:20}
\end{eqnarray}
Only particles passing through all the apertures can get to the PB origin to redefine the number of particles, the direction, and the mean square angle as
\begin{eqnarray}
n_b = {n_0}_b \, T_b,\qquad T_b = \int\hspace{-0.8em}\int_{-\pi}^{\pi} \frac{ \rmd\theta \, \rmd\phi}{2\pi{\sigma_\theta}_b^2} \, \rme^{-\frac{\theta^2+\phi^2}{2{\sigma_\theta}_b^2}} \hspace{-2.5em} \prod_{\qquad \{a \mid z_a > z_0\}} \hspace{-2.5em} {T_a}_{i'_a j'_a}, \label{eq:21}
\\
\left(\frac{{v_b}_x}{{v_b}_z},\frac{{v_b}_y}{{v_b}_z}\right) = \frac{1}{T_b} \int\hspace{-0.8em}\int_{-\pi}^{\pi} \frac{\rmd\theta\,\rmd\phi}{2\pi{\sigma_\theta}_b^2} \rme^{-\frac{\theta^2+\phi^2}{2{\sigma_\theta}_b^2}} \hspace{-2.5em} \prod_{\qquad \{a \mid z_a > z_0\}} \hspace{-2.5em} {T_a}_{i'_a j'_a} \left(\frac{{{v_p}_x}}{{{v_p}_z}}, \frac{{{v_p}_y}}{{{v_p}_z}}\right), \label{eq:22}
\\
\overline{\theta^2}_b = \frac{1}{T_b} \int\hspace{-0.8em}\int_{-\pi}^{\pi} \frac{\rmd\theta\,\rmd\phi}{2\pi{\sigma_\theta}_b^2} \rme^{-\frac{\theta^2+\phi^2}{2{\sigma_\theta}_b^2}} \hspace{-2em} \prod_{\qquad\{a \mid z_a > z_0\}} \hspace{-2em} {T_a}_{i'_a j'_a} \frac{\theta^2+\phi^2}{2}, \label{eq:23}
\end{eqnarray}
where $i'_a j'_a$ is the aperture pixel in which line $(\theta,\phi)$ intersects.
In practice, these integrals are made numerically at $0.2\,{\sigma_\theta}_b$ sampling intervals for $\pm 3\,{\sigma_\theta}_b$ regions.

\subsubsection{Range compensation}
The RCF shortens the residual range of the pencil beam by the thickness of the intersecting pixel as $\Delta R_b = -{S_{\rm C}}_{i_{\rm C} j_{\rm C}}$ and increases the mean square angle by $\Delta \overline{\theta^2}_b$ in \eref{eq:6} before the beam is transported downstream.

\subsubsection{Downstream collimation}

Every pencil beam is individually transported by \eref{eq:5}--\eref{eq:7} through downstream apertures.
At an aperture, which is practically either the top or bottom face of an optional PTC, pencil beams near the edge will be partially transmitted.
Incidentally, edge distance ${d_a}_{i_a j_a}$ in \eref{eq:4} naturally corresponds to density-interface distance $d_{\rm int}$ in \eref{eq:8} for PB splitting.
At every downstream aperture, multiplicity $M$ is appropriately determined while limiting overreaching particles to below 2\% by setting $\kappa_{\rm n} = 3$.
In the case of splitting, daughter beams are defined according to \eref{eq:9}--\eref{eq:12} and then individually transported downstream starting from the current aperture with possible recursive splitting in the same manner.
The pencil beams that are finally out of the aperture will be blocked by setting $n_b = 0$, which addresses the partial-blocking effect of the collimator.

\subsection{Experimental validation}

\subsubsection{Apparatus}

\begin{figure}
\begin{indented}
\item[] \includegraphics[width=10 cm]{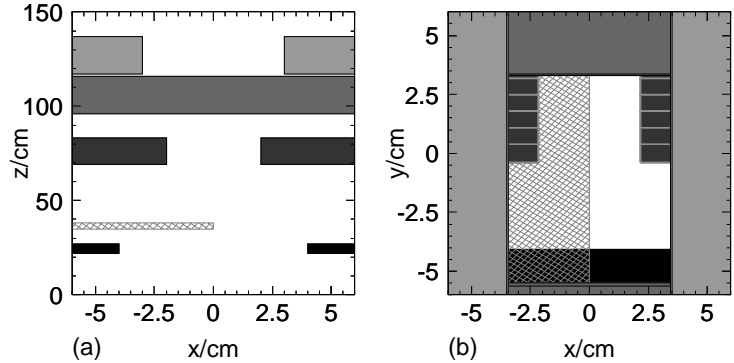}
\end{indented}
\caption{Beam-customization devices of the experiment; (a) the $y=0$ cross-section view, (b) the beam's eye view on the isocenter plane, where the filled areas represent XJC, YJC, MLC, and PTC from upstream to downstream and the hatched area represents RCF.}
\label{fig:2}
\end{figure}

As this study shares the objective of beam-customization modeling with the former study in the PB-redefinition approach (Kanematsu \etal 2008b), we use the same experimental data, where a broad carbon-ion beam of residual range $R_0 = 19.6$ cm in water was customized with an XJC at height 117--137 cm, a YJC at 96--116 cm, and a partially effective MLC at 69--83 cm, a 3-cm PMMA half-plate RCF at 35--38 cm, and an 8-cm-square PTC at 22--27 cm as shown in \fref{fig:2}. 
Four lines of the in-air dose profiles on the isocenter plane were measured along the $x$ axis at $y = -2$ cm and $1$ cm and along the $y$ axis at $x = -1$ cm and $1$ cm.
The 20\%--80\% penumbra sizes ($\approx 1.68\,\sigma_t$) were 0.58 cm for the XJC edge and 0.48 cm for the YJC edge, which translate into rms source sizes $\sigma_{{\rm S}x} = (0.58/1.68) (940-117)/117 = 2.43$ cm at $z_{{\rm S}x} = 940$ cm and $\sigma_{{\rm S}y} = (0.48/1.68)(1040-96)/96 = 2.81$ cm at $z_{{\rm S}y} = 1040$ cm.
The tissue-air ratio for the 3-cm PMMA ($\rho = 1.16$) was measured to be 0.951.

\subsubsection{Implementation}

In the calculation, $100 \times 100$ dose grids in a single layer were arranged on the isocenter plane at 1-mm intervals.
The open field of uniform fluence ($\Phi_0 = 1$) was subdivided into the BEV image pixels of size $\delta = 0.5$ mm on the isocenter plane, to each of which a pencil beam was defined at the effective scattering point of the RCF.
For every pencil beam, upstream collimation by the XJC, the YJC, and the MLC, range shift and scattering by the RCF, and beam transport including collimation and splitting by the PTC down to the isocenter plane were applied. 
The in-air dose distribution on the isocenter plane was calculated with
\begin{eqnarray}
\fl
D(x,y) = \sum_{b} \frac{n_b\, {D_\Phi}_b}{2\pi\,{\sigma_t}_b^2}\, \rme^{-
\frac{\left(x-x_b\right)^2 + \left(y-y_b\right)^2} {2\,{\sigma_t}_b^2}},
\quad
{D_\Phi}_b = \cases{1 & for $R_b \approx 19.6$ cm\\ 0.951 & for $R_b \approx 16.1$ cm,} \label{eq:24}
\end{eqnarray}
where ${D_\Phi}_b$ is the tissue-air ratio or the dose per fluence for beam $b$.

To verify the effectiveness of PB splitting for the PTC edge, we calculated dose distributions at heights 0 cm (isocenter plane) and 20 cm (immediate downstream) by relocating the dose grids and compared them with corresponding non-splitting calculations, for which we disabled splitting by setting $\kappa_{\rm d} = 0$.

\section{Results}

\begin{figure}
\begin{indented}
\item[] \includegraphics[width=8.5 cm]{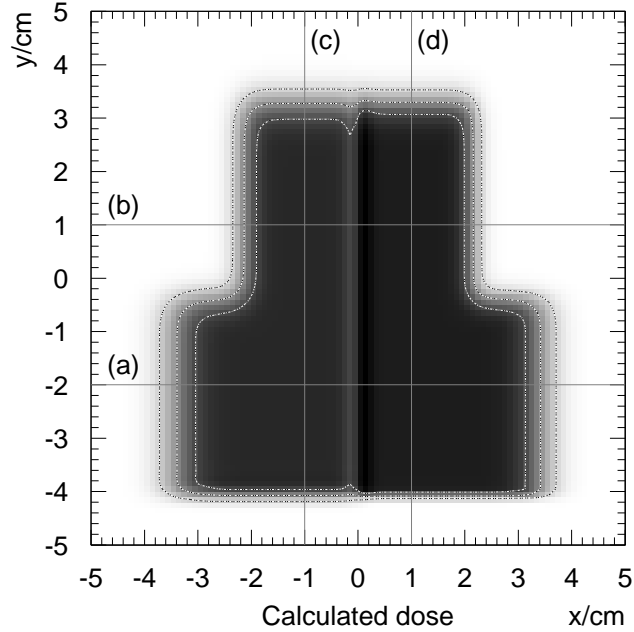}
\end{indented}
\caption{Calculated dose distribution in gray scale for the customized carbon-ion field, where the dotted contours represent 20\%, 50\%, and 80\% dose levels and the gray lines indicate the dose-profiling positions.}
\label{fig:3}
\end{figure}

In the calculation, 40000 beams were originally defined at the RCF, 23912 of them passed through the upstream collimators, and 20444 of them passed through the PTC to end up with 36704 dose-contributing beams by splitting.
The CPU time of a 2.4 GHz Intel Core 2 Duo processor amounted to 1.30 s and 1.25 s for the calculations with and without PB splitting.
\Fref{fig:3} shows the calculated dose distribution. 
The dip and bump along the $y$ axis are attributed to scattering by the PMMA half plate.
Sharpness of the field edge was strongly correlated with the distance to the effective collimator.

\begin{figure}
\begin{indented}
\item[] \includegraphics[width=8.5 cm]{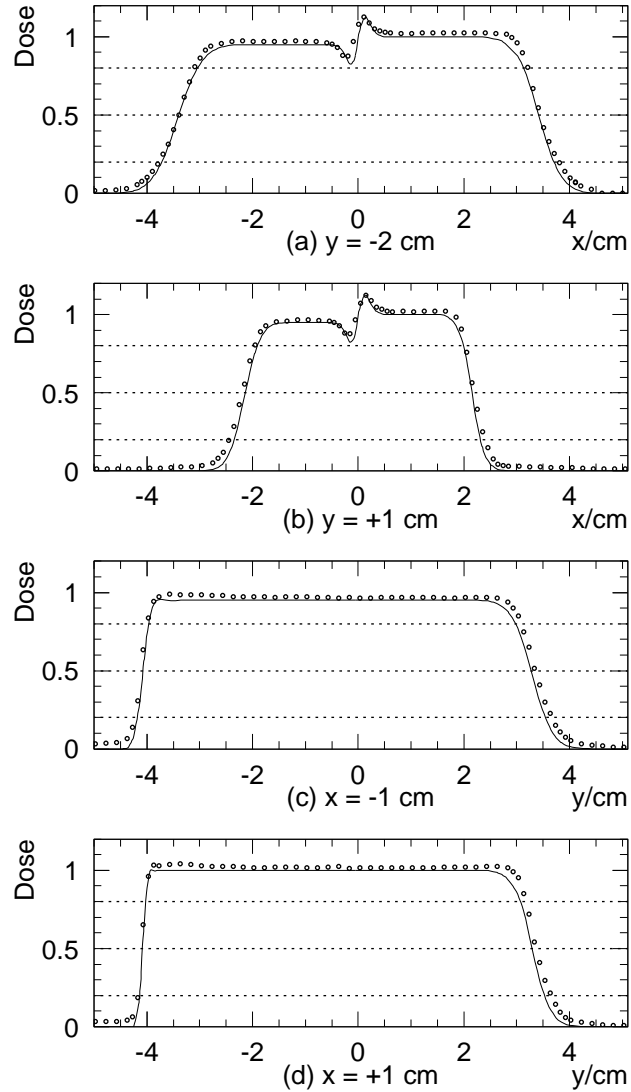}
\end{indented}
\caption{Dose profiles along the $x$ axis at $y = -2$ cm (a), $y = +1$ cm (b), along the $y$ axis at $x = -2$ cm (c), and $x = +1$ cm (d), where the solid lines are the calculations and the open circles are the measurements.}
\label{fig:4}
\end{figure}

In the experiment, the uncertainty of the scanned detector positions was 0.1~mm and that of the collimator positions was $\lesssim 0.5$~mm according to the specifications.
The latter may only shift the edge position and will not influence the penumbra size.
The single-point dose uncertainty was evaluated to be 0.3\% in repeated measurements, which is negligible for penumbra analysis.
\Fref{fig:4} shows the calculated and measured doses profiles, where the measured doses are in fact the dose ratios of the customized field to the open field to compensate for the fluence non-uniformity.
Unexpectedly, the customized-field doses were higher than the open-field doses by a few percent.
That may be attributed to the contribution of particles hard-scattered by the collimators, which was not considered in the present model.

From these profiles, the 20\%--80\% penumbra sizes were obtained by reading 20\% and 80\% dose positions by linear interpolation of two sampling points, which brings dominant uncertainty amounting to a fraction of the sampling interval of 1 mm.
The measured penumbra sizes were then corrected to quadratically exclude $1.68\, \sigma_{\rm size}$ with effective dosimeter size $\sigma_{\rm size} = 0.5$ mm for a 2-mm$\phi$ pinpoint chamber.
\Tref{tab:1} summarizes the resultant penumbra sizes. 
These measurements and calculations agreed to a submillimeter level, which is consistent with the estimated uncertainty.

\begin{table}
\caption{Measured and calculated 20\%--80\% penumbra sizes for the customized carbon-ion field.}
\begin{indented}
\item[]
\begin{tabular}{l l l l l }
\br
Profiling & Interested- & Effective & \multicolumn{2}{l}{Penumbra size (mm)} \\
\cline{4-5}
position & edge side & device(s) & measurement & calculation \\
\mr
$y=-2$ cm & $x$ left & XJC+RCF & 6.4 & 6.7 \\
$y=-2$ cm & $x$ right & XJC & 5.8 & 5.8 \\
$y=+1$ cm & $x$ left & MLC+RCF & 4.6 & 4.4 \\
$y=+1$ cm & $x$ right & MLC & 3.7 & 3.2 \\
$x=-1$ cm & $y$ lower & PTC+RCF & 2.3 & 2.6 \\
$x=-1$ cm & $y$ upper & YJC+RCF & 5.6 & 5.7 \\
$x=+1$ cm & $y$ lower & PTC & 1.4 & 1.3 \\
$x=+1$ cm & $y$ upper & YJC & 4.8 & 4.5 \\
\br
\end{tabular}
\end{indented}
\label{tab:1}
\end{table}

\begin{figure}
\begin{indented}
\item[] \includegraphics[width=8.5 cm]{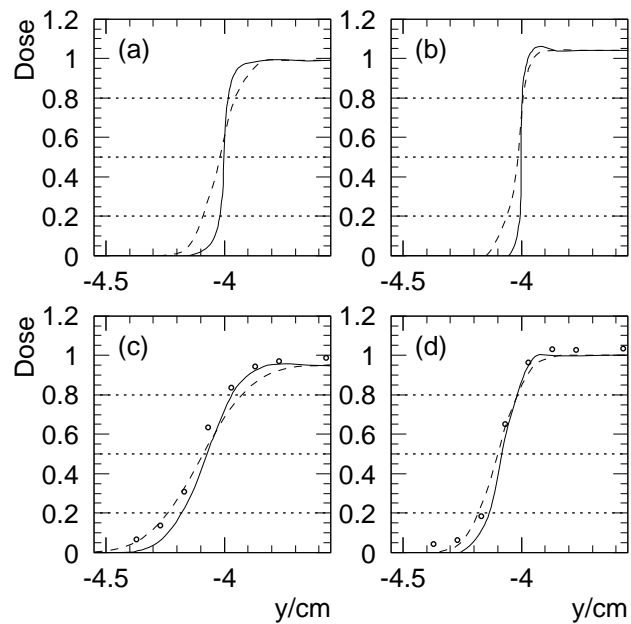}
\end{indented}
\caption{Dose profiles for PTC edge; (a) immediate downstream with RCF, (b) immediate downstream without RCF, (c) isocenter plane with RCF, (d) isocenter plane without RCF, where the solid and the dashed lines are calculations with and without PB splitting, respectively, and the open circles are measurements.}
\label{fig:5}
\end{figure}

The effectiveness of beam splitting for the PTC-edge sharpening is shown in \fref{fig:5}, where panels (c) and (d) show enlarged views of  panels (c) and (d) in \fref{fig:4} with additional lines for non-splitting calculations.
The PB splitting reasonably sharpened the field edges at the immediate downstream and
made better agreement with the measurements on the isocenter plane.
Ironically, the contamination of collimator-scattered particles happened to compensate substantially for the lack of edge sharpening in the tail regions.

\section{Discussion}

It is one of the algorithmic novelties of this study to originate the pencil beams at the effective scattering points of the RCF regardless of upstream collimation.
Then, the lateral heterogeneity of the RCF is naturally irrelevant to the minimized pencil beams.
Upstream collimation is reasonably modeled as filtering of particles in the angular distribution to correct the phase-space parameters of the defined pencil beams.

While the PB size and density are generally arbitrary in PB algorithms, small size and high density are required to represent sharp edges of downstream collimation.
In the present model, the sharp PTC edge was naturally realized by splitting of the pencil beams.
In the former study (Kanematsu \etal 2008b), because the pencil beams could not be redefined as monochromatic after range compensation, they were only artificially downsized for edge sharpening.
The downsizing strength was empirically determined to reproduce the 20\%--80\% penumbra size on the isocenter plane while overlooking the other aspects.
In fact, while the resultant penumbra sizes were equivalently good for both models, the dose profiles in the upstream was unphysically bouncy in the former study due to insufficient density, which could be clinically problematic.

In the original PB-splitting algorithm (Kanematsu \etal 2009), the overreaching condition was defined as the one-standard-deviation distance ($\kappa_{\rm d} = 1$) to a 10\% density change.
That was because its objective heterogeneity was moderate density variation among body tissues.
This study deals with solid and precisely defined collimator edges, for which the $\kappa_{\rm d} = 3$ distance to an aperture edge may be more appropriate.

In the present example, the PTC was effective only for approximately 1/4 of the field edge.
The PB splitting was limited to the pencil beams around the effective edge and actually increased the number of beams by 80\% and the CPU time by 4\%.
This discrepancy is mainly attributed to computational overhead for generation, upstream collimation, and range compensation of the pencil beams.
Although we only dealt with the planer grids in this case, the PB splitting would not add severe computational load even for volumetric grids when used with the grid-dose-spreading convolution (Kanematsu 2011).

In heavy charged particle radiotherapy, target doses are predominantly formed by Bragg peaks of primary particles.
Hard-scattered particles are generally out of the scope of practical PB algorithms due to difficulty in their modeling.
Fortunately, the collimator-scattered particles tend to lose large energy in the collimator and thus naturally attenuate with depth (van Luijk \etal 2001).
Nevertheless, Kimstrand \etal (2008) included the collimator-scatter contribution in a convolution algorithm using Monte-Carlo-generated kernels.
That approach may be valid and will further improve the accuracy if combined with the present model.

\section{Conclusions}

We have developed a calculation model for customization of a broad beam of heavy charged particles based on the PB-splitting algorithm.
In this model, a broad beam is decomposed into pencil beams of various size that is necessarily and sufficiently small to deal with structures of the beam-customization devices accurately and efficiently.
Also, placement of the PB origins at the effective scattering points in the RCF effectively reduced the relevant heterogeneity and greatly simplified the algorithm using only monochromatic pencil beams.

The performance of the model was tested against existing experimental data, which demonstrated that the penumbra size for various collimator edges in a single field was accurate to a submillimeter level.
This beam-customization part can be naturally combined with the patient-dose-calculation part that is similarly based on the PB-splitting algorithm (Kanematsu 2011) to complete an accurate and efficient dose calculation algorithm for treatment planning of heavy-charged-particle radiotherapy.

\References

\item[] Borgefors G 1986 Distance transformations in digital images {\it Comput. Vision Graph. Image Process.} {\bf 34} 344--371

\item[] Coutrakon G, Bauman M, Lesyna D, Miller D, Nusbaum J, Slater J, Johanning, J and Miranda, J and DeLuca Jr P M and Siebers J 1991 A prototype beam delivery system for the proton medical accelerator at Loma Linda {\it Med. Phys.} {\bf 18} 1093--9

\item[] Eyges L 1948 Multiple scattering with energy loss {\it Phys. Rev.} {\bf 74} 1534--5

\item[] Goitein M 1978 A technique for calculating the influence of thin inhomogeneities on charged particle beams {\it Med. Phys.} {\bf 5} 258--264

\item[] Gottschalk B, Koehler A M, Schneider R J, Sisterson J M and Wagner M S 1993 Multiple Coulomb scattering of 160 MeV protons {\it Nucl. Instrum. Methods B} {\bf 74} 467--90

\item[] Gottschalk B 2010 On the scattering power of radiotherapy protons {\it Med. Phys.} {\bf 37} 352--367

\item[] Hong L, Goitein M, Bucciolini M, Comiskey R, Gottschalk B, Rosenthal S, Serago C and Urie M 1996 A pencil beam algorithm for proton dose calculations \PMB {\bf 41} 1305--30

\item[] Kanai T \etal 1999 Biophysical characteristics of HIMAC clinical irradiation system for heavy-ion radiation therapy {\it Int. J. Radiat. Oncol. Biol. Phys.} {\bf 44} 201--10

\item[] Kanematsu N, Matsufuji N, Kohno R, Minohara S and Kanai T 2003 A CT calibration method based on the polybinary tissue model for radiotherapy treatment planning \PMB {\bf 48} 1053--64

\item[] Kanematsu N, Akagi T, Takatani Y, Yonai S, Sakamoto H and Yamashita H 2006 Extended collimator model for pencil-beam dose calculation in proton radiotherapy \PMB {\bf 51} 4807--17

\item[] Kanematsu N, Yonai S and Ishizaki A 2008a The grid-dose-spreading algorithm for dose distribution calculation in heavy charged particle radiotherapy {\it Med. Phys.} {\bf 35} 602--7

\item[] Kanematsu N, Yonai S, Ishizaki A and Torikoshi M 2008b Computational modeling of beam-customization devices for heavy-charged-particle radiotherapy \PMB {\bf 53} 3113--27

\item[] Kanematsu N 2009 Semi-empirical formulation of multiple scattering for the Gaussian beam model of heavy charged particles stopping in tissue-like matter \PMB {\bf 54} N67--73

\item[] Kanematsu N, Komori M, Yonai S and Ishizaki A 2009 Dynamic splitting of Gaussian pencil beams in heterogeneity-correction algorithms for radiotherapy with heavy charged particles \PMB {\bf 54} 2015--27

\item[] Kanematsu N 2011 Dose calculation algorithm of fast fine-heterogeneity correction for heavy charged particle radiotherapy {\it Physica Medica} (in press) doi:10.1016/j.ejmp.2010.05.001

\item[] Kimstrand P, Traneus E, Ahnesjo A and Tilly N 2008 Parametrization and application of scatter kernels for modelling scanned proton beam collimator scatter dose \PMB {\bf 53} 3405--29

\item[] Kohno R, Kanematsu N, Kanai T and Yusa K 2004 Evaluation of a pencil beam algorithm for therapeutic carbon ion beam in presence of bolus {\it Med. Phys.} {\bf 31} 2249--53

\item[] Lambert J, Suchowerska N, McKenzie D R and Jackson M 2005 Intrafractional motion during proton beam scanning \PMB {\bf 50} 4853--62

\item[] Li Y, Zhang X, Dong Lei and Mohan R 2007 A novel patch-field design using an optimized grid filter for passively scattered proton beams \PMB {\bf 52} N265--ZN275

\item[] Petti P L 1992 Differential-pencil-beam dose calculations for charged particles {\it Med. Phys.} {\bf 19} 137--49

\item[] Shiu A S and Hogstrom K R 1991 Pencil-beam redefinition algorithm for electron dose distributions {\it Med. Phys.} {\bf 18} 7--18

\item[] van Luijk P, van’t Veld A A, Zelle H D and Schippers J M 2001 Collimator scatter and 2D dosimetry in small proton beams {\PMB} {\bf 46} 653--67

\endrefs

\end{document}